\documentclass[aps,prf,twocolumn,reprint,superscriptaddress]{revtex4-1}

\usepackage{graphicx}
\usepackage{anysize}
\usepackage{lineno}
\usepackage[amssymb]{SIunits}
\usepackage{lipsum}
\usepackage{amsmath}
\usepackage{amssymb}
\usepackage[toc,page]{appendix}
\usepackage{SIunits}
	\usepackage{soul}
\usepackage{color}
\usepackage{bm}

\newcommand \be {\begin{equation}}
\newcommand \ee {\end{equation}}
\newcommand \bea {\begin{eqnarray}}
\newcommand \eea {\end{eqnarray}}

\newcommand{\mbf}[1]{\mathbf{#1}}

\newcommand{\subfigletter}[4]{%
\begin{tikzpicture}%
    \node[anchor=south west,inner sep=0] (a) at (0,0) {#1};%
    \node[above left = #4 and #3 of a] {#2} ; 
\end{tikzpicture}%
}%

\usepackage{tikz} 
\usetikzlibrary{positioning}
\usepackage{xcolor}
\definecolor{colorred}{HTML}{e41a1c}
\definecolor{colorblue}{HTML}{377eb8}
\definecolor{colorgreen}{HTML}{4daf4a}  

\begin{document}


\title{Deflection of phototactic microswimmers through  obstacle arrays}

\author{Marvin Brun-Cosme-Bruny}
\email[]{These authors contributed equally to this work.}
\affiliation{Universit\'e Grenoble Alpes, CNRS, LIPhy, F-38000 Grenoble, France}
\author{Andre F\"ortsch}
\email[]{These authors contributed equally to this work.}
\affiliation{Theoretische Physik, Universit\"at Bayreuth, D-95447 Bayreuth, Germany}
\author{Walter Zimmermann}
\email[]{walter.zimmermann@uni-bayreuth.de}
\affiliation{Theoretische Physik, Universit\"at Bayreuth, D-95447 Bayreuth, Germany}
\author{Eric Bertin}
\email[]{eric.bertin@univ-grenoble-alpes.fr}
\affiliation{Universit\'e Grenoble Alpes, CNRS, LIPhy, F-38000 Grenoble, France}
\author{Philippe Peyla}
\email[]{philippe.peyla@univ-grenoble-alpes.fr}
\affiliation{Universit\'e Grenoble Alpes, CNRS, LIPhy, F-38000 Grenoble, France}
\author{Salima Rafa\"{\i}}
\email[]{salima.rafai@univ-grenoble-alpes.fr}
\affiliation{Universit\'e Grenoble Alpes, CNRS, LIPhy, F-38000 Grenoble, France}

\date{\today}

\begin{abstract}
We study  the effect of  inhomogeneous  environments on the swimming direction 
of the microalgae \textit{Chlamydomonas Reinhardtii} (CR)  
in the presence of a light stimulus. Positive or negative phototaxis describe the ability of microorganisms 
to bias their swimming towards or away from a light source. 
Here we consider microswimmers with negative phototaxis in a microfluidic device with a microfabricated 
square lattice of pillars as obstacles. We measured a mean deflection of 
 microswimmers that shows an interesting nonlinear dependence on the direction of
the guiding light beam with respect to  the symmetry axes of
the pillar lattice. 
By simulating a model swimmer in a pillar lattice and analyzing its scattering behavior, we 
identified the width of the  reorientation distribution of swimmers to be also crucial for 
the nonlinear behavior of the swimmer deflection.  
On the basis of  these  results we  suggest in addition 
an analytical model  for microswimmers, where the pillar lattice is replaced by an  anisotropic scattering medium,  that depends only 
on a scattering rate and the width of the reorientation distribution of swimmers.  This flexible and handy model
fits the experimental results as well.  
The presented analysis of the deflection of light guided swimmers  through  pillar lattice may be used for
separating swimmers having different reorientation distributions.
\end{abstract}

\maketitle

\section{Introduction}
Far from any walls, planktonic micro-organisms swim freely, while in a complex environment 
they often adhere or attach on surfaces \cite{an1998concise},
for example, when bacterial colonies are embedded in biosynthesized extracellular 
polymeric substances \cite{costerton1999bacterial, lecuyer2015focus}. This can lead to the so-called 
bio-fouling \cite{ping2005membrane}. In microfluidic devices \cite{marty2012formation}, 
the same process can occur and leads to a destruction of the device. 
The way micro-organisms move inside and colonize a porous medium such as a membrane or a filter 
is a subject of current research \cite{rusconi2011secondary, creppy2019effect, volpe2011microswimmers}.  The statistics of transport of microswimmers through a crowded environnement have been explored in recent works \cite{Chepizhko2019, Maas2019}. Collective patterns like vortices have also been reported for swimming bacteria in arrays of pillars \cite{Sokolov2018}.

Can microswimmers be guided through 
complex environments?  Deformable particles  such as (red blood) cells or even simple dumbbells driven by a fluid flow 
through inhomogeneous landscapes show interesting deflection scenarios \cite{Austin:2004.1,Bammert:2009.1,Bridle:2014.1}.
For instance,   particle loaded flows through arrays of pillars  are 
 a very important  microfluidic technique that enables a continuous size- or deformability dependent particle
sorting with exceptional resolution, depending on the relative orientation 
between the flow direction and a symmetry axis of obstacle arrays \cite{Austin:2004.1,Bridle:2014.1}. 
In the case of the  phototactic microswimmer   {\it Chlamydomonas Reinhardtii} 
(CR), the  position of a light source   defines instead of the flow  a preferred direction \cite{garcia2013light}.
This light orientation of the algae CR is rather effective and
leads for instance  to self-focusing to jets of microswimmers in Poiseuille flow  \cite{Peyla:2010.1} including
interesting jet instabilities \cite{Peyla:2014.1, lauga2017clustering}. 
Investigations of the interaction of self-propelled particles
with a complex environment in general is a challenging current research topic with various applications  
\cite{Chepizhko2013, Bechinger2016, chamolly2017active, Bertrand2018, Morin2017, Zeitz2017, alonso2019transport}.

\newcommand{\arrlen}{1.cm}
\begin{figure}[!b]
\centering
    \hspace*{-.2cm}\begin{tikzpicture}
        \node[anchor=south west,inner sep=0] at (2cm,0) {\includegraphics[width=\dimexpr \columnwidth - 2cm \relax]{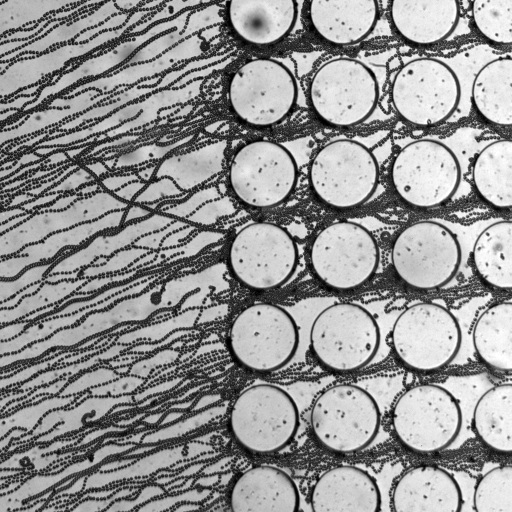}};
        \node[] at (1.cm,5.0cm)  {\rotatebox{22}{\textcolor{colorblue}{Light Source}}} ; 
        \foreach \x in {1.1,3.6,...,9.1} { \draw[->,color=colorblue,>=stealth, line width=2mm] (0.4cm,\x*0.5cm) --++ (22:1.52cm); } 
        \draw[line width=0.5mm] (0.4cm,0.5cm) --++(\arrlen,0); 
        \draw[line width=0.5mm] (\dimexpr 0.4cm + \arrlen \relax,0.47cm) node[above right]{$\theta_\ell$} arc (0:23:\arrlen); 
        \draw[fill=white, draw=black, line width=0.2mm] (3.85cm,0.47cm) rectangle ++(0.4cm,0.5cm); 
        \draw[line width=0.5mm,color=white] (3.8cm,0.47cm) node[above right]{\textcolor{black}{$\theta_{\rm i}$}} arc (0:23:\arrlen)  ; 
        \draw[line width=0.5mm,color=white] (3.8cm,0.5cm) --++(-\arrlen,0) --++ (23:\arrlen); 
        \draw[fill=white, draw=black, line width=0.2mm] (6.cm,0.57cm) rectangle ++(0.4cm,0.5cm); 
        \draw[line width=0.5mm,color=white] (5.95cm,0.57cm) node[above right]{\textcolor{black}{$\theta_{\rm f}$}} arc (0:15:\arrlen)  ; 
        \draw[line width=0.5mm,color=white] (5.95cm,0.6cm) --++(-\arrlen,0) --++ (15:\arrlen); 
        \draw[|-|,line width=0.5mm] (6.545cm,5.5cm) --++(0.38cm,0); 
        \draw[fill=white, draw=black, line width=0.2mm] (6.35cm,5.2cm) rectangle ++(1,-0.5);
        \node[color=black] at (6.85,4.9) {$100\micro\meter$};
    \end{tikzpicture}
  \caption{
  Trajectories of phototactic microalgae
CR through a  microfluidic device. 
  A LED light-beam source is used with a tunable incidence angle ${\theta}_\ell$  with respect to the horizontal 
  $x$-axis of the square lattice of pillars. CR respond to the light stimulus through negative 
  phototaxis and flee from the light source. In the pillar-free region, their swimming  direction $\theta_{\rm i}$ is
  on average directed along the light, i.e. ${\bar \theta}_{\rm i} \sim \theta_\ell$. After entering the pillar lattice, 
  successive reorientations of swimmers cause  a distribution of swimmer trajectories with orientations   $\theta_{\rm f}$  
  and an average swim orientation $\bar{\theta}_{\rm f}$.  }  \label{schema_photo_carre}
\end{figure}
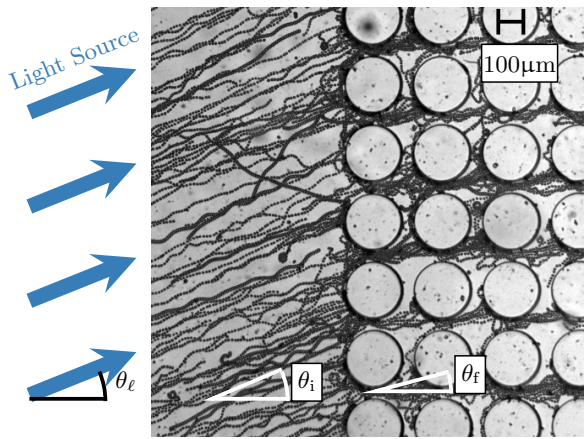
 
Here, we investigate negative phototactic motile algae CR  
moving away from a light source through a microfluidic device with a transparent and regular lattice
of microfabricated pillars as described in Sec.~\ref{setup}.
These motile algae experience by the pillar lattice a deviation between their average swim directions
and the light beam, cf.  Sec.~\ref{expres}.
In order to understand the origin of such deviations, %
 we perform simulations taking only into account collisions between swimmers and pillars 
and statistical reorientations. By comparing these results with Lattice Boltzmann (LB) simulations 
that include hydrodynamic interactions (HI), we can extract the role of different key features, 
as described in Sec. \ref{sec:numerical}.

The numerical simulations can account for our experimental results whereby the intrinsic orientational noise of the swimmers is crucial 
for  a broadening of the distribution of the swim orientation around the  light beam. 
In Sec.~\ref{sec:analytical}  a simple analytical model is developed, which is also closely connected to the numerical analysis. Both approaches cover the essential experimental observations. A discussion of several results and conclusions is given in Sec. ~\ref{sec:conclus}.


\section{Experimental set-up \label{setup}}

We use as a microswimmer model the green micro-alga
CR, a biflagellate photosynthetic and phototactic cell of $\unit{10}{\micro\meter}$ 
diameter \cite{harris2009chlamydomonas}. The microalgae are grown under a 14h/10h light/dark cycle at $22\degree$C 
and are harvested in the middle of the exponential growth phase. 
CR's front flagella beat in a breast stroke manner and propel the microswimmer in the fluid \cite{Jibuti2017}. 
The swimming motion is characterized by a persistent random walk 
in the absence of a bias \cite{polin2009chlamydomonas,garcia2011random}. 
However, in the presence of a light stimulus (green wavelength, i.e., around 510 nm), 
microalgae tend to swim away from the light source  \cite{garcia2013light}.
Suspensions are used at an initial volume fraction of about 0.05\%, 
so that the HI among microswimmers is negligible. The cells are finally introduced within a chamber 
containing a square lattice of $\unit{200}{\micro\meter}$-diameter pillars regularly spaced 
by  a minimal surface-to-surface distance $d=\unit{30}{\micro\meter}$. 
The pillar lattice has been designed such that the length of a unit cell of pillars is comparable to the persistence length of swimmers, to allow for the reorientation of swimmers while passing through it.
Pillars are made of transparent PDMS by means of soft lithography processes \cite{qin2010soft}. 
Both the diameter and inter-pillar distance are kept constant. The height of 
pillars is $\unit{70}{\micro\meter}$  corresponding to about 7 cell diameters. 
Bovine Serum Albumine is used to coat the pillars in order to limit 
adsorption of cells. The space surrounding the complex environment is free of pillars. \\

We observe the cells under a bright field microscope. 
We use an inverted microscope (Olympus IX71) coupled with a CCD camera (AVT GX3300) 
used at a frame rate of 15 fps. Using a low magnification objective ($\times 2$) 
allows us to get a wide field of view ($\unit{3614}\times\unit{2885}{\micro\meter}^2$) 
to be able to acquire both the pillar-free region and complex medium at the same time. 
The sample is enclosed in an covering box with two red filtered windows for visualization. 
This prevents the microscope light from triggering phototaxis.\\

 At the beginning microswimmers are homogeneously distributed in the chamber. 
 A white LED light is switched on with a tunable orientation angle $\theta_{\rm \ell}$ 
 with respect to the horizontal axis of the square lattice of pillars,
 as shown in Fig.~\ref{schema_photo_carre}. Due to negative phototaxis, microswimmers move away from 
 the light source and go through the lattice of pillars as depicted in Fig.~\ref{schema_photo_carre}.


 \section{Experimental results\label{expres}}

Particle tracking is performed with the library Trackpy \cite{trackpy, crocker}. Orientations of 
microswimmers can then be extracted as the mean orientation of a trajectory over 0.5 seconds. 
\begin{figure}[!htb]
\centering
  \includegraphics[width=8 cm]{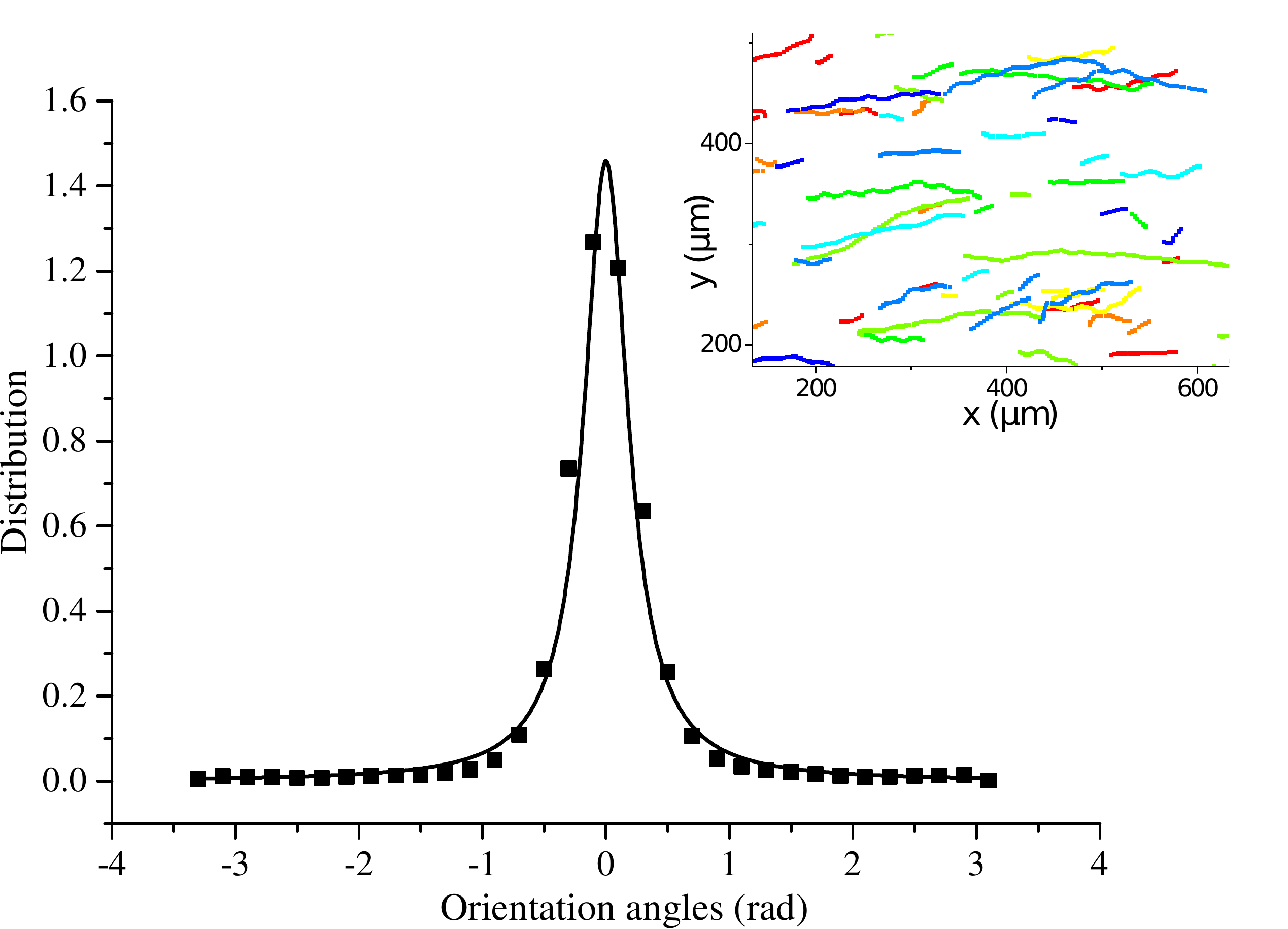}  	   
  \caption{The distribution of CR swimming directions in a pillar-free medium in 
  the presence of a light source positioned at the left side with an incidence angle $\theta_\ell=0$. 
  The inset shows  swimmer trajectories. The solid line is the distribution given by Eq.~(\ref{Lorentz}).}  \label{orientation_simplemedium}
\end{figure}
\begin{figure*}[htb]
\centering
  \hspace*{0cm}
  \includegraphics[width=14.5cm]{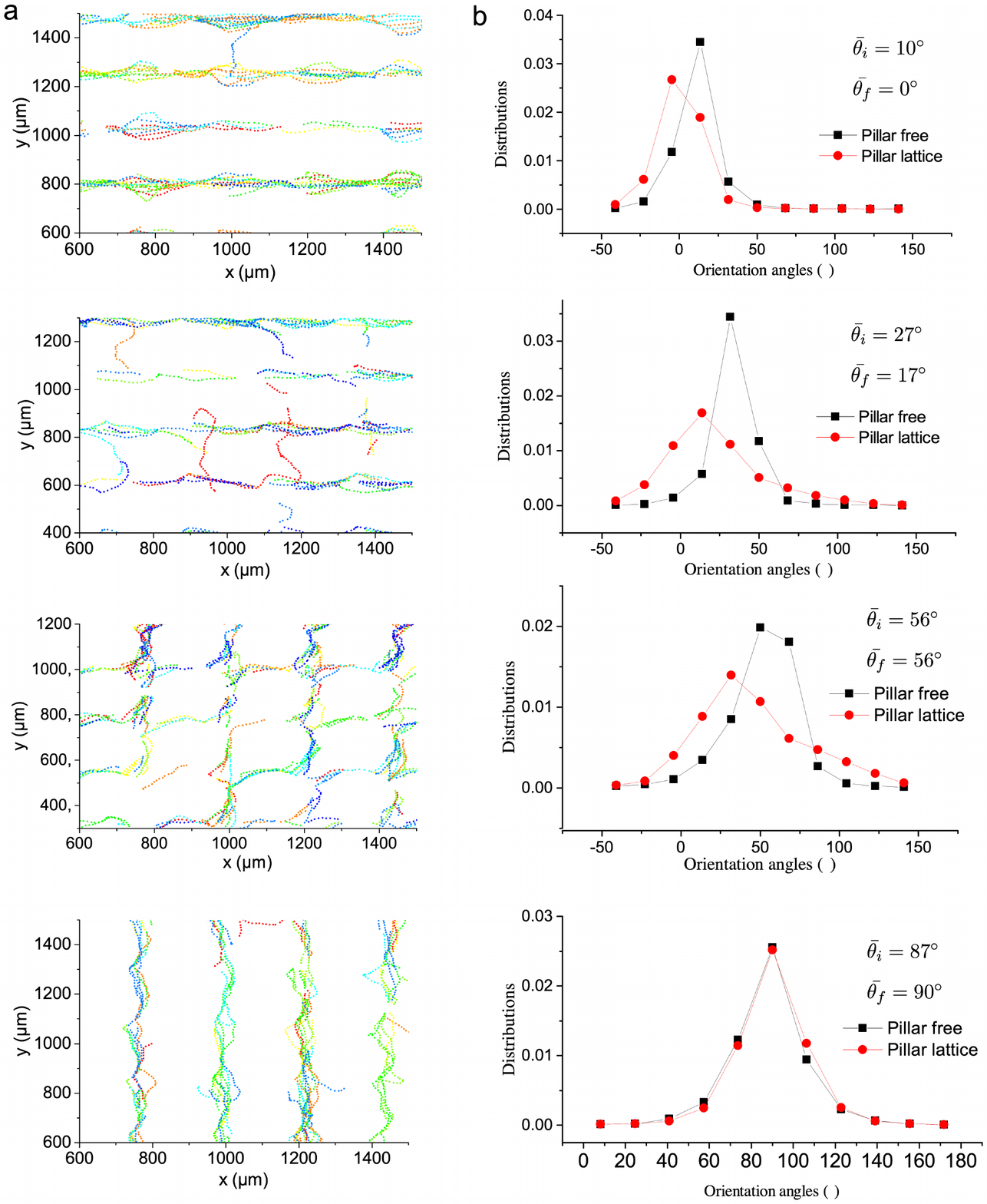}  
  \vspace{-2cm}
  \caption{The left column (part a)) shows
  for four examples of  different incident angles $\theta_\ell \sim \bar{\theta}_{\rm i}$
   experimentally tracked trajectories of CR swimmers through the pillar lattice.
  For the same four angles, the right column (part b)) shows the distributions of the CR-trajectory 
  orientations through the pillar lattice,
  described by $\theta_{\rm f}$, and they are compared with the distribution of  CR-trajectory orientation $\theta_{\rm i}$
  in the pillar free range.
  }
  \label{trajectoires_photo_carre}
\end{figure*}

Fig. \ref{orientation_simplemedium} shows the distribution of  the orientation of microswimmer trajectories
in a pillar-free medium, as well as examples of  swimmer trajectories in the inset. In the pillar-free region,
the average swimming direction ${\bar \theta}_{\rm i}$ corresponds to the orientation $ \theta_\ell$   of the light beam. The maximum  
of the orientational distribution around the direction of the light beam is found to be close
to a truncated Lorentzian distribution as previously shown in Ref.~\cite{martin2016photofocusing},
\begin{equation}
\label{Lorentz}
\Psi(\theta)=\frac{\Gamma}{2\pi}\frac{1}{\frac{\Gamma^2}{4}+\theta^2}\,,
\end{equation}
where we obtain $\Gamma=0.436\,{\rm rad}=25\degree$ for the full width at half maximum.
 
\begin{figure}[!tb] 
	\centering
	\includegraphics[width=\columnwidth]{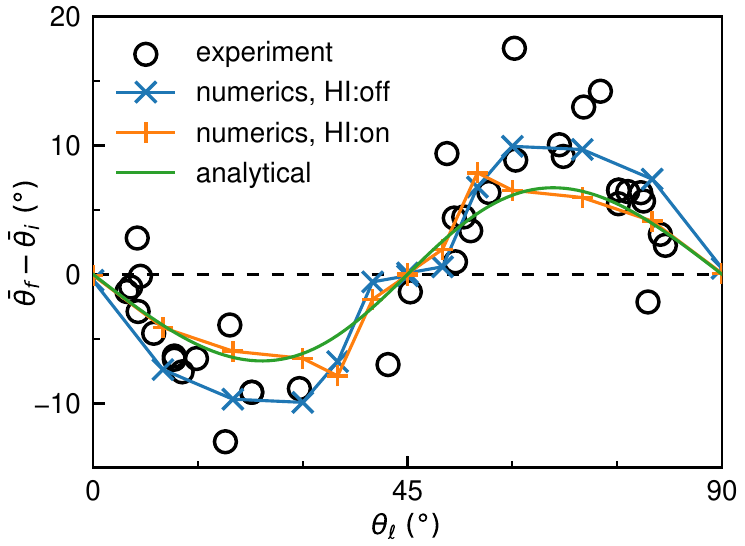} 
	\caption{ 
	The difference $\bar{\theta}_{\rm f}-  \bar{\theta}_{\rm i}$ between the
	mean swimming direction $\bar{\theta}_{\rm f}$ through the pillar lattice is shown as a function 
	of the  mean orientation $\bar{\theta}_{\rm i}$  through the pillar free range.
	The experimental data are given by black circles. The numerical results from Sec.~\ref{sec:numerical} without hydrodynamic interactions (HI) 
	are given by the blue symbols and with HI  by the red symbols (for $\Gamma=25\degree$).
	 The green solid curve shows the result for the analytical model from Sec.~\ref{sec:analytical} for
	    $A_3=0.174$ from Eq.~(\ref{A3}) with 
	$D_{\rm R}=0.37s^{-1}$, $\alpha=0.33s^{-1}$ and $\lambda_0=2.09s^{-1}$.
	}  
	\label{angle_experimental}
\end{figure}

As shown in Fig.~\ref{trajectoires_photo_carre} the distribution of the swimmer-orientation angles $\theta_{\rm f}$ within the pillar lattice 
and therefore 
the mean swimming-direction $\bar{\theta}_{\rm f}$ depend on the angle $\theta_\ell$ of the light beam.
We show for 
 four different values of $\theta_\ell \sim \bar{\theta}_{\rm i}=10^o,~27^o,~56^o,~87^o$ in part a) examples of trajectories 
 of the CR swimmer. In part b) we show  for these four angles simultaneously the distribution of  $\theta_{\rm f}$ within the pillar lattice
 and $\theta_{\rm i}$ in the pillar free range.
The distributions of  $\theta_{\rm f}$ are found to be narrower
when the incident beam of light is oriented towards
the lattice axes  ${\theta}_\ell \approx 0\degree$ (or ${\theta}_\ell \approx 90\degree$).  
In these cases the trajectories through the pillar lattice follow the directions of corridors 
aligned with the light direction.
 For other values of
${\theta}_\ell$ the angular distribution of 
$\theta_{\rm f}$ is broader since CR are scattered by the pillar lattice. 
Therefore, we find in the pillar region
$\bar{\theta}_{\rm f}\approx {\theta}_\ell$ when ${\theta}_\ell \approx 0\degree$ or  ${\theta}_\ell \approx 90 \degree$.
On the contrary,
for other angular values,  the mean orientation $\bar{\theta}_{\rm f}$ 
deviates from $\bar{\theta}_{\rm i}$ since CR are scattered by the pillars away from the light direction.  
The maximum deviation occurs around
${\theta}_\ell \approx 30\degree$ where we find  $\bar{\theta}_{\rm f} \approx 10\degree$, as can be seen in 
Fig. \ref{angle_experimental}. Note that the curve is symmetric with respect to ${\theta}_\ell=45\degree$, 
where we find  $\bar{\theta}_{\rm f} \approx {\theta}_\ell$. 
By comparing our experimental results with numerical simulations
we would like to understand on the one hand the role --if any-- of HI and on the other hand the role
of the intrinsic noise on the reorientations of the cells away from the light direction, 
i.e., the full width $\Gamma$ of the distribution $\Psi(\theta)$ in 
 Fig.\ref{orientation_simplemedium}.


\section{Numerical swimmer model}
\label{sec:numerical}

To further understand the observed deflection behavior shown in Fig.~\ref{angle_experimental} we complement in this section
our experimental results by a numerical analysis of a swimmer model of CR introduced in
Sec.~\ref{swimodel} and described in the appendix. An analysis of the swimmer trajectories and their orientational distributions
provides a basic picture of swimmer deflection and a thorough foundation of the analytical model given in Sec.~\ref{sec:analytical}.

\subsection{Swimmer model\label{swimodel}}

\begin{figure}[!ht]
  \begin{center}
\includegraphics[width=6cm]{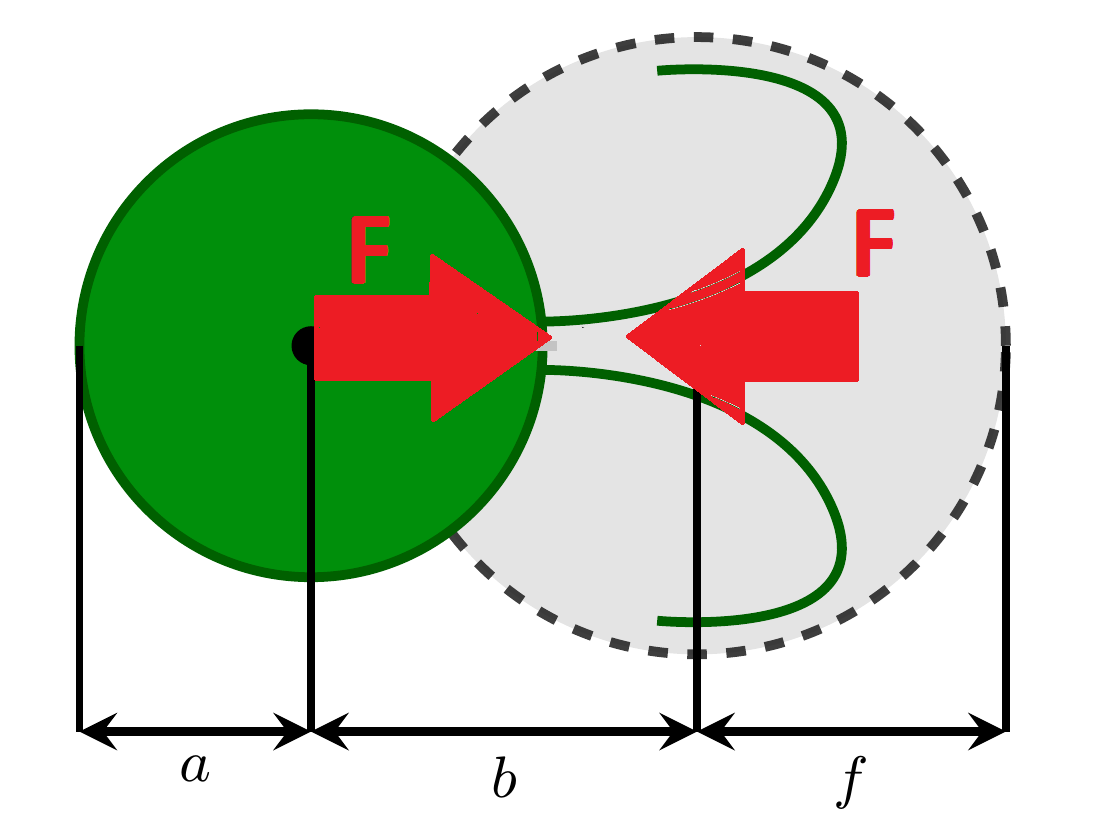}\\  
\caption{Sketch of a model of the swimmer CR as used in simulations.
The swimmer, described in more detail in Appendix~\ref{appswimmmodel}, consists of  a hard impenetrable sphere of radius $a$ (green).
It is complemented by a sphere of  radius $f=\frac{4}{3} a$ at a distance $\frac{5}{3}a$ from the body center, that covers the range
of flagella motion: It is permeable for the fluid but hardcore repulsive for objects like other swimmers. A doublet of forces is exerted by the swimmer on the fluid.}
\label{fig:sim_model}
 \end{center}
\end{figure}

For our numerical analysis we introduce  a force dipole 
model for  CR algae as illustrated 
in Fig.~\ref{fig:sim_model}.
The spherical body of radius $a$ is impenetrable for the fluid and experiences a drag during its
motion through the fluid. The flagella are located in a region of radius $\frac{4}{3}\,a$ with 
a distance $\frac{5}{3}\,a$ to the body-center. This region --unlike the body-- is permeable for the fluid, but is taken 
into account for hard core interactions with other swimmers or obstacles and mimics the excluded volume shape for the region 
covered by the flagella  motion \cite{Kantsler:2013,Lushi:2017,Schwarzendahl:2018,Contino:2015}. %
A doublet of forces is applied to the fluid both by flagella and the body (Fig.~\ref{fig:sim_model}). 
The resulting flow-profile is shown in Fig.~\ref{fig:sim_model_app} in the appendix and resembles 
the experimentally observed averaged flow profile of a CR algae \cite{Drescher:2010} moving at a velocity $V_0$.

For the equations of motion described in Appendix~\ref{appendi2} we use a 3D-Lattice Boltzmann (LB) solver \cite{chen1998lattice} that covers 
the full hydrodynamics between swimmers and obstacles or walls. We also use a dissipative collision model to test 
the effect of pure collisional interactions between motile particles and the pillar wall without the influence of HI.
In both cases,  the phototaxis is modeled as a preferential direction of motion: each swimmer is, after an 
exponentially distributed time of mean $\tau_{\rm ph}$, reoriented towards a direction $\theta$ randomly drawn from a Lorentzian distribution of
mean $\theta_\ell$ (restricted to $-\pi < \theta<\pi$, with $\theta=0$ corresponding to the $x$-axis) , that reproduces the
truncated Lorentzian distribution shown in Fig.~\ref{orientation_simplemedium}. The average time $\tau_{\rm ph}$ is chosen as $\approx \frac{70a}{V_0}$, close to
the experimental value ($\unit{2}{\second}$).

The model swimmer is immersed in a simulation cuboid domain with in-plane periodic boundary conditions 
and with a single cylindrical pillar placed in the middle of the domain, which reproduces the exact proportions 
of the experiment.  %

\subsection{Numerical deflection of trajectories}
For the numerical deflection data, we place a swimmer with random initial position and direction in the simulation region. 
We simulate the trajectories for different initial conditions for each value of $\theta_\ell$.
From the averaged swimming direction,  we extract the deflection angle ${\bar \theta}_{\rm f}-\theta_\ell$. 
This is repeated for different light beam angles $0<\theta_\ell<\pi/4$. For the data in the range of $\pi/4< \theta_\ell < \pi/2$ we generated the data from simulations in the range $0<\theta_\ell<\pi/4$ by using the point symmetry of the system. 

The swimmer trajectories are simulated either with the LB method, which takes the hydrodynamic interactions between 
the swimmer and the pillar walls into account or by the
dissipative collision model (DCM), also described in Appendix~\ref{appendi2}. The dependence of the deflection angle, i.e.,  deviation
$\bar{\theta}_{\rm f}- \bar{\theta}_{\rm i}$ from the light beam orientation $\theta_\ell\sim \bar{\theta}_{\rm i}$ is shown in Fig. \ref{angle_experimental} 
together with the experimental data. Surprisingly, the results of both simulation approaches
fit the characteristics of the experimental data quite well. Therefore, HI is not crucial on a qualitative level  
for the deflection process. 
It turns out that the occurrence of the deflection is mainly influenced by geometric properties and the statistical distribution 
of the reorientation.
To get a better understanding of the underlying processes, we perform in the next section 
a statistical analysis 
of the scattering of a swimmer on a single pillar lattice without hydrodynamics and without a light beam. Later we extend 
the results on the scattering mechanism to full trajectories in the presence of light.

\subsection{Deterministic scattering without light\label{Scattstat}}
In order to reach a basic understanding of the swimmer scattering leading to deflections as in Fig.~\ref{angle_experimental}
and to provide a foundation of the  anisotropic analytical model in
Sec.~\ref{sec:analytical}, we analyze the deterministic trajectories of a model 
swimmer during a single scattering process (through a single pillar unit cell) in the absence of light. 
\begin{figure}[!tbp]
    \includegraphics[width=7.5cm]{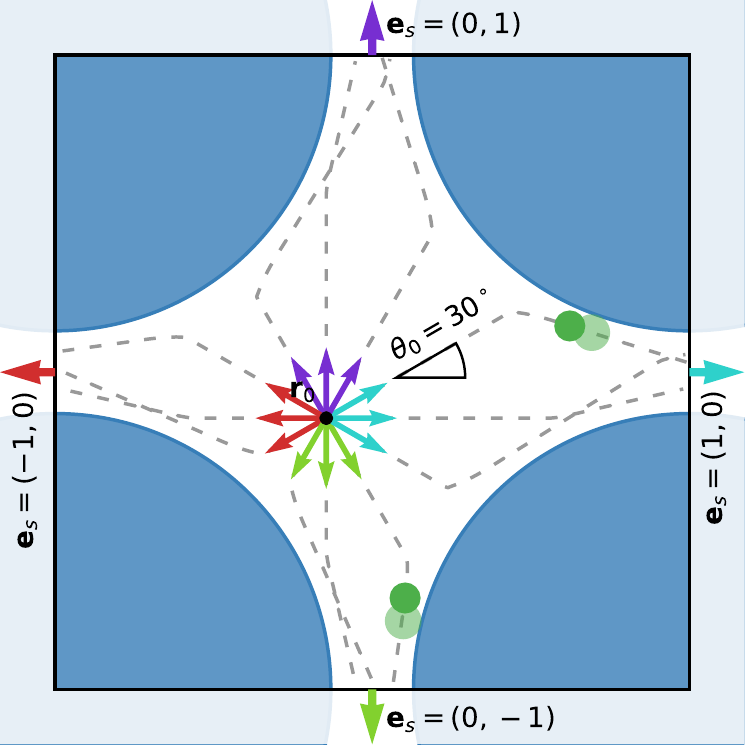}
    \caption{Sketch of the swimmer scattering by pillars.
    A swimmer starts at the position $\mathbf{r}_0$ with 
    initial angle $\theta_0$, corresponding to an initial direction $\mathbf{e}_{\rm i}$. Here we have sketched $12$ different directions $\mathbf{e}_{\rm i}$ at the same ${\bf r}_0$.
    The swimmer trajectories (gray dashed lines) are then determined via  DCM and swimmers
    (sketched as green circles) may be deflected by the pillars. 
    We track the trajectories until a swimmer leaves the unit cell at one of the exits.  The exit directions are $\mathbf{e}_{\rm s}(\mathbf{r}_0,\theta_0)=\{(\pm 1,0),(0,\pm1)\}$ (cyan, red, violet, green arrows). 
    A statistics on the exit vectors is obtained by repeating the simulation for many different  $\mathbf{r}_0$ and $\theta_0$.
}
    \label{fig:numstat_sketch}
\end{figure}
With this aim in mind, we place  the model swimmer at different initial positions $\mathbf{r}_0$ and with different
initial directions $\mathbf{e}_{\rm i}$ within a unit of the pillar free space. 
The initial angle enclosed by  $\mathbf{e}_{\rm i}$ and the $x$-axis is  $\theta_0$. 
We then determine the swimmer trajectory with the DCM. On their path the swimmers are scattered at the pillars
due to the excluded volume effects. They leave a pillar unit cell through one of the four 
exits between the pillars,
as shown in Fig.~\ref{fig:numstat_sketch}. 
This procedure is repeated many times for different, uniformly distributed initial positions $\mathbf{r}_0$ and angles $\theta_0$ 
to obtain the directions $\mathbf{e}_{\rm s}(\mathbf{r}_0,\theta_0)$ at the exit
as a function of the initial position and orientation. 
The four exit directions are $ \mathbf{e}_{\rm s}(\mathbf{r}_0,\theta_0)=\{(\pm 1,0),(0,\pm1)\}$,
and the angle of the exit vector towards the positive x-axis 
is defined as $\theta_{\rm s}$ ($\theta_{\rm s}=\{0\degree,180\degree,90\degree,-90\degree\}$).
\begin{figure}[!tbp]
     \includegraphics[width=7.5cm]{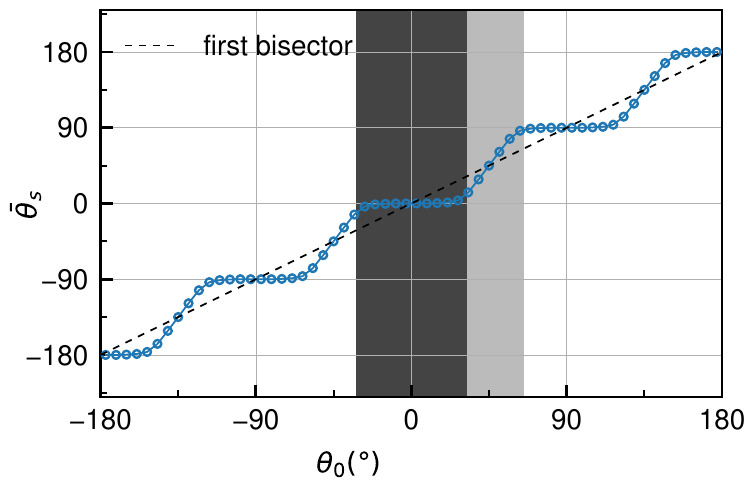}
     \caption{Angle $\bar{\theta}_{\rm s}$ after scattering in a single pillar cell, averaged over 
     all initial positions ${\bf r}_0$ shown as a function of the initial  swimmer orientation $\theta_0$. 
     For initial angles $\theta_0$ around $0\degree,~\pm 90\degree,~180\degree$ the swimmers are deflected by pillars such that they  
      are channeled 
     along the symmetry axes of the pillar-lattice, which is the origin of the plateaus of $\bar{\theta}_{\rm s}$.
     For initial angles $\theta_0 \sim \pm 45\degree, \pm 135\degree$ swimmers are equally likely 
     deflected into neighboring exits, which results in $\bar{\theta}_{\rm s} \sim \pm 45\degree, \pm 135\degree$.}
     \label{fig:numstat_theta_s}
\end{figure}
Fig.~\ref{fig:numstat_theta_s} shows the  scattering (or exit) angle $\bar{\theta}_{\rm s}$ averaged over the equally distributed initial 
positions ${\bf r}_0$ as a function of
the initial orientation $\theta_0$.
Swimmers with ${\bf r}_0$ near the center between the four pillars 
and an initial orientation $\theta_0\sim 0$ are very likely to leave the unit cell via the right exit  $\mathbf{e}_{\rm s}=(1,0)$.
  With our simulations we find for initial angles in the range  $0\degree<\theta_0\lesssim 30\degree$
  that  swimmers are channeled by collisions with the pillars to the right exit $ \mathbf{e}_{\rm s}=(1,0)$ as well.
 In this range of $\theta_0$  the mean scattering angle $\bar{\theta}_{\rm s}(\theta_0)\sim 0$ is nearly constant 
as indicated by the dark area in Fig.~\ref{fig:numstat_theta_s}. 
If the initial orientation $\theta_0$ is  increased then, depending on the initial position ${\bf r}_0$, the
swimmers are deflected with increasing probability to the upper exit direction $ \mathbf{e}_{\rm s}=(0,1)$. 
With an initial orientation $\theta_0 = 45\degree$ swimmers starting at different initial positions ${\bf r}_0$ 
are  deflected in average equally likely either to the exit $ \mathbf{e}_{\rm s}=(1,0)$ or to the exit $ \mathbf{e}_{\rm s}=(0,1)$, which results
in an  average exit angle  $\bar{\theta}_{\rm s} \sim 45\degree$, cf. Fig.~\ref{fig:numstat_theta_s}.
With $\theta_0 \in [60\degree,120\degree]$ we find  $\bar{\theta}_{\rm s} \sim 90\degree$. A similar behavior 
is found around $\theta=180\degree$ and $\theta=-90\degree$ as indicated in Fig.~\ref{fig:numstat_theta_s} as well.

\begin{figure}[!tbp]
     \includegraphics[width=7.5cm]{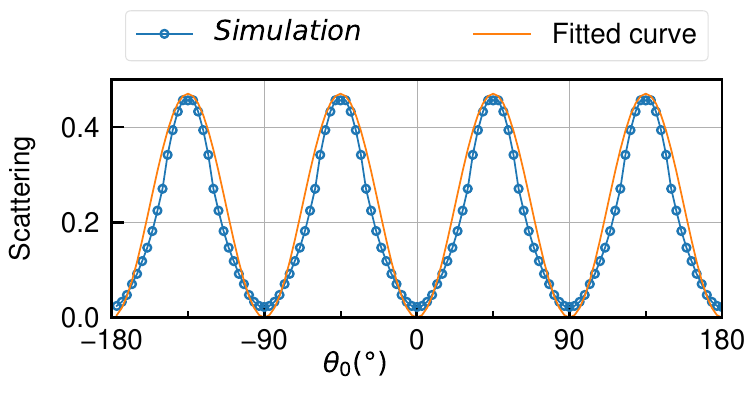}
     \caption{The blue line shows the scattering function $\left<1-\mathbf{e}_{\rm i} \cdot \mathbf{e}_{\rm s} \right>_{\mathbf{r}_0}$
     for the scattering in a single pillar cell, averaged over the initial positions 
      ${\bf r}_0$  as a function of initial direction ${\bf e}_{\rm i}$ resp.     
      the initial angle $\theta_0$. 
     Along the pillar free symmetry axes the scattering function vanishes 
     and takes its maxima along the directions in between the symmetry axes.
     The orange curve is a fit by the expression $\lambda_0 \left[ 1 - \cos(4\theta_0)\right]$
     to these exemplary numerical scattering results with $\lambda_0=0.235$.
      } 
      
     \label{fig:numstat_scatter} 
\end{figure}

Swimmers scattered by crossing 
a unit cell can be described by the following scattering function averaged over all
initial positions in a unit cell: $\left< 1-\mathbf{e}_{\rm i} \cdot \mathbf{e}_{\rm s}\right>_{\mathbf{r}_0}$. 
The averaged function is 
shown in Fig.~\ref{fig:numstat_scatter} as a function of the initial orientation $\theta_0$.
The scattering function almost vanishes with initial swimmer orientations close  to one of the pillar-free axis and it has maxima along the
'diagonal' directions $\theta_0 \approx  \pm45\degree,  \pm135\degree$. 
The scattering function has a period of four in the range $[-\pi<\theta_0<\pi]$,
which reflects the symmetry axes of the pillar lattice. 
In addition, these numerical results for the scattering of a single swimmer provide a 'microscopic' foundation for the assumption of a scattering rate $\lambda_0 \left[ 1 - \cos(4\theta_0)\right]$
made in Eq.~(\ref{eq:Ltheta0}) of our phenomenological anisotropic 
scattering model, see Sec.~\ref{sec:analytical} below. 
This form of the scattering rate reflects both the four-fold symmetry of the pillar lattice and the fact that swimmers are not scattered with their mean swim direction along pillar symmetry axes.

\subsection{Deflection in the presence of light}
\begin{figure}[!htbp]
    \includegraphics[width=7.5cm]{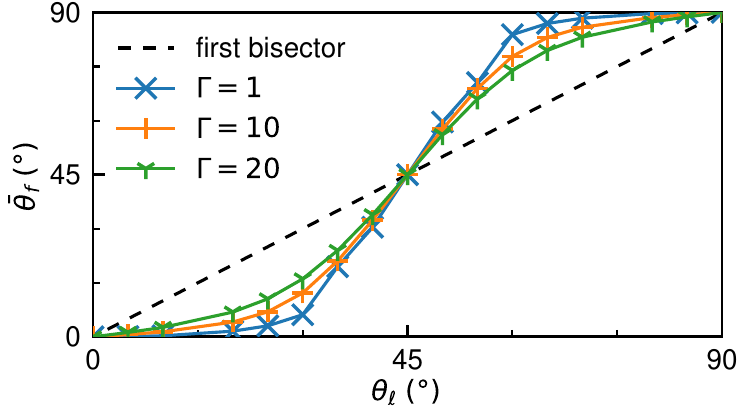}
    \caption{The average deflection angle $\bar{\theta}_{\rm f}$ extracted from the single scattering function is shown as a function of the light orientation $\theta_\ell$ 
    for three values of $\Gamma$ corresponding to three different width of the Lorentz distribution, see Eq.~(\ref{Lorentz}).}
    \label{fig:numstat_lightdeflection}
\end{figure}

The trajectory of CR can be described by a run-and-tumble walk with a preferred direction in the presence of light. 
That means a swimmer reorients after a certain time towards a new direction, 
loosing all information about the previous direction and path. 
Swimmers reorient on average every $\sim \unit{2}{\second}$ towards the light orientation. 
With a speed of $\unit{100}{\micro\meter/\second}$ and a length of the unit cell of $L=\unit{200}{\micro \meter}$,
a swimmer reorients on average at least once when crossing a pillar unit cell. 
Because it looses information about the past after this reorientation, the swimmer trajectories 
through several pillars can be described by repeated single scattering processes in a single pillar unit cell,
provided one uses periodic boundary conditions.
In the previous section we determined the function that gives the swimmers direction $\bar{\theta}_{\rm s}(\theta_0)$ 
after a single scattering as a function of the initial direction $\theta_0$. The initial direction of a scattering 
process is the direction after the tumbling. Since we know the Lorentz probability distribution of swimmer reorientations
given by Eq.~(\ref{Lorentz}), we can extract the mean swimming angle $\bar{\theta}_{\rm f}$ 
of a swimmer's full trajectory by using the scattering function in a single unit cell. 
For this we need to weight the occurrence of the directions after scattering $\mathbf{e}_{\rm s}(\theta_0)$ 
according to the tumbling probability distribution $\psi(\theta_0-\theta_\ell)$ and get the mean swimming direction
\begin{align}
 \bar{\mathbf{e}}_{\rm f} = \left< ~ \bar{\mathbf{e}}_{\rm s}(\theta_0) \psi(\theta_0-\theta_\ell) ~\right>_{\theta_0} ,
\end{align}
with $\bar{\mathbf{e}}_{\rm s}=\left< \mathbf{e}_{\rm s} \right>_{\mathbf{r}_0}$ and $\theta_{\rm f}$ as the angle between $\bar{\mathbf{e}}_{\rm f}$ and the $x$-axes.
Note that the norms of $\bar{\mathbf{e}}_{\rm s}$ and $\bar{\mathbf{e}}_{\rm f}$ are not necessarily equal to one.
For very narrow distributions (i.e., small $\Gamma$) $\psi$ is approximately a delta distribution. 
That means for the mean swimming direction $ \bar{\mathbf{e}}_{\rm f} \approx \left< \bar{\mathbf{e}}_{\rm s}(\theta_0) \delta(\theta_0 -\theta_\ell) \right>_{\theta_0}  = \bar{\mathbf{e}}_{\rm s} (\theta_\ell)$, 
so the angle $\bar{\theta}_{\rm f}$ approaches the scattering function from the previous section. 
This case is shown in Fig.~\ref{fig:numstat_lightdeflection} (blue curve). For this small value of $\Gamma$, 
the mean swimming as a function of the light angle $\theta_\ell$ has a rather steep behavior around
$\theta_\ell \sim 45^\degree$ and
with a channeled regime $\bar{\theta}_{\rm f}$ for $\theta_\ell \lesssim 30\degree$ 
similar to Fig.~\ref{fig:numstat_scatter} in the range $0<\theta_0<90\degree$. 
If we increase the distribution width $\Gamma$ we see in Fig.~\ref{fig:numstat_lightdeflection} 
that the mean swimming direction $\bar{\theta}_{\rm f}$ changes from the steplike, 
channeled function towards the first bisector. This behavior is caused by the  distribution $\psi$. 
If $\Gamma$ is small, the initial directions $\mathbf{e}_{\rm s}$ nearly always point towards the light direction. 
Therefore,  if the light is along a symmetry axis, the  scattering of the swimmers is with a 
high probability such that they are  channeled through the pillar lattice along a symmetry axis.
If we choose a broad reorientation distribution width $\Gamma$, 
even for a light orientation and the initial orientations  ${\bf e}_{\rm i}$ close to a symmetry axis   it becomes with increasing $\Gamma$
more and more probable that swimmers are scattered away from the respective symmetry axis. Thus the mean swimming direction
tends towards the first bisector for large $\Gamma$.

This is a crucial insight: the width of the Lorentz distribution determines the steepness of the shape of $\bar{\theta}_{\rm f}(\theta_\ell)$.
This is rather independent of whether the width of the distribution
is just an intrinsic property of swimmers or possibly caused by other effects, such 
hydrodynamics interactions between several swimmers as discussed  in  Sec.~\ref{sec:hydro}.
Furthermore we obtain a profound insight on the swimming statistics from the single-scattering function $\theta_{\rm s}(\theta_0)$. 
This technique could probably be adapted to other problems, without necessity of simulating the swimmer trajectories but using information of absolute value of $\bar{\mathbf{e}}_s$, adapted weighting of the scattering function (e.g., position dependent), multiple folding for temporal correlation etc., to investigate the effect of depletion zones (e.g., caused by an imposed fluid flow), temporal correlation or different geometries of obstacle placement and many more.

\subsection{Effect of the hydrodynamic interaction}
\label{sec:hydro}
To identify the influence of hydrodynamic interactions (HI), we reduce the noise due to the tumbling in this section.
This is achieved by simulations with a small distribution width of $\Gamma=1^\circ$ and a reduced tumbling time of $\tau_{\rm ph} \approx 8a / V_0$.

\begin{figure}[h!tbp] 
    \centering
  \subfigletter{\includegraphics[width= 3.333cm ]{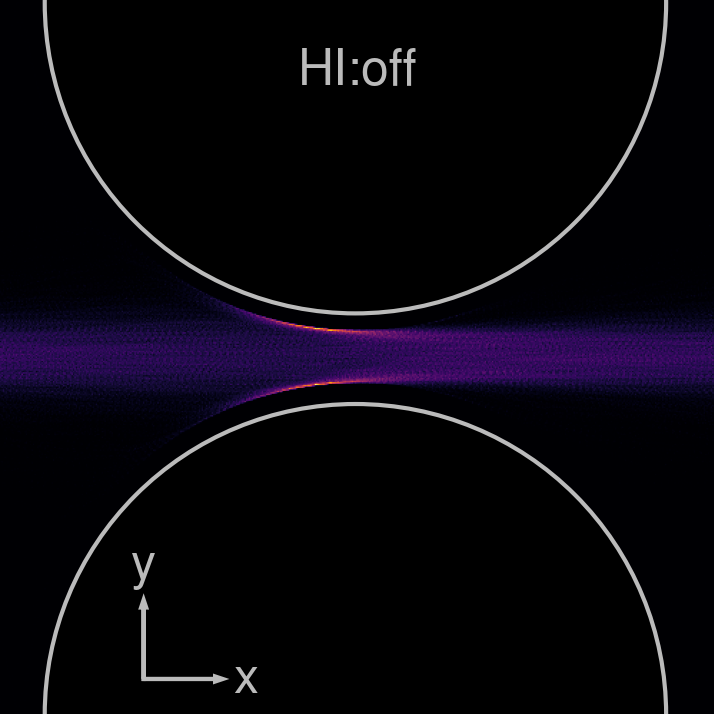}}{a) $\mathcal{P}_{DCM}$}{-2.65cm}{0.cm}~%
  \subfigletter{\includegraphics[width=4.1666cm]{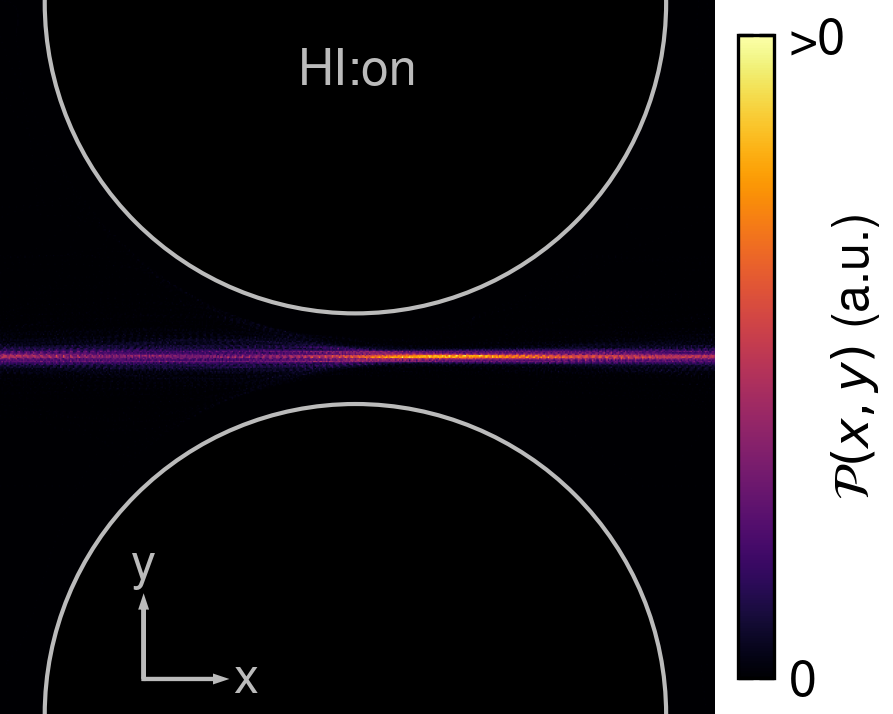}}{b) $\mathcal{P}_{LB}$}{-2.65cm}{0.cm}
  \caption{Probability distribution $\mathcal{P}(x,y)$ of the position of
  a single swimmer in a unit cell  for $\Gamma=1$, $\tau_{\rm ph}\approx 8a/V_0$ and $\theta_\ell=0$. 
  a) Without hydrodynamic interactions (DCM simulations), the interaction between swimmers and pillars is collision based,
  leading to a high probability to find particles at the contact positions near 
  the pillars. b) For simulations with HI (LB simulations), we find pronounced focusing between the pillars. This indicates that hydrodynamics help the swimmers to avoid collisions with the obstacles.
  }
  \label{fig:position_probability}
\end{figure}

The effects of hydrodynamic interactions (HI) becomes important for the interaction between swimmers and pillars.
We show in Fig.~\ref{fig:position_probability} 
the probability distribution $\mathcal{P}(x,y)$ 
of the position of a  guided single swimmer for   $\theta_\ell=0$.
In part a) the probability distribution  $\mathcal{P}(x,y)$ is shown for the case without HI  between a swimmer and 
the pillars. This distribution is considerably broader than in part b) where in LB simulations
the HI between pillars and swimmers is taken into account.

\begin{figure}[h!tbp]
 \includegraphics[width=\columnwidth]{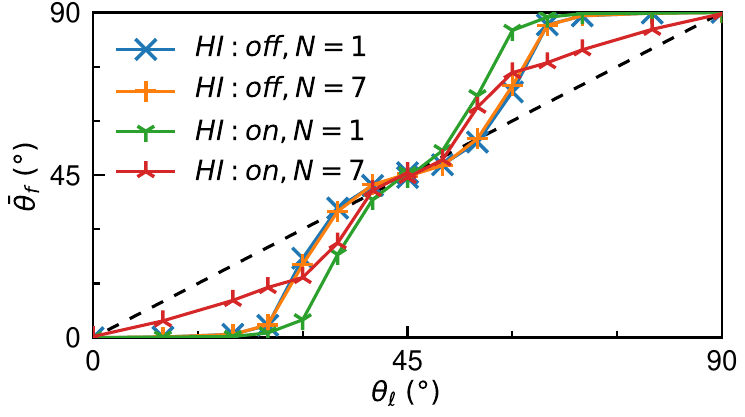}
 \caption{Deflection curves for simulations with different numbers ($N=1,7$) of swimmers in a unit cell of the
 pillar lattice for $\Gamma=1$ and $\tau_{\rm ph}\approx 8a/V_0$. For the simulations without HI, there is no difference for one or seven 
 swimmers in a unit cell. In the case with HI in LB simulations  the deflection curve 
 is far less step for seven swimmers than for one swimmer (see text). }
 \label{fig:deflection_numcomp}
\end{figure}

The effect of an enhanced swimmer channeling via HI is also confirmed by the
deflection curves $\bar{\theta}_{\rm f}(\theta_\ell)$ for single swimmers in Fig.~\ref{fig:deflection_numcomp}. 
The simulation of a single swimmer  with HI (green curve) show
up to an angle $\bar{\theta}_\ell\sim 30\degree$  a channeling behavior while  swimmers without HI (blue curve)
 escape channeling already at about $\bar{\theta}_{\rm f}\approx 20\degree$. 

The influence  of hydrodynamic interactions becomes also
significant in the case with several swimmers in a unit cell. 
In Fig.~\ref{fig:deflection_numcomp} we compare the mean deflection $\bar{\theta}_{\rm f}$ obtained by simulations 
for a single swimmer without
HI (blue curve), with the deflection of a single swimmer out of seven swimmers (orange) without HI.
There is no significant difference.

When HI are taken into account the 
situation for a single swimmer and seven swimmers in a unit cell is rather different. 
As mentioned, the channeling of a swimmer is stronger in the case with than without HI. 
However, the deflection curve $\bar{\theta}_{\rm f}(\theta_\ell)$ of a single swimmer out of seven
swimmers is in the case with HI less steep than for a single swimmer with HI. Moreover, 
it is also less steep 
than for swimmers without HI. 

This can be explained as follows: The hydrodynamic interaction between different swimmers is of nonlinear nature.
It is well known  that nonlinear interactions between several particles cause a more complex dynamics than
one obtains for  a single particle. Hence, the HI between the
swimmers act like an additional external noise source on a single swimmer. This additional noise does not depend on the orientation of the light 
nor on the intrinsic  random reorientations swimmers. 
However, the hydrodynamic interactions between the swimmers 
cause additional reorientations of the single swimmer. The HI driven additional reorientations 
have a similar effect as a broader Lorentz distribution on a single swimmer. 
Therefore a boarder reorientation Lorentz distribution has on a single swimmer a similar effect as the hydrodynamic interactions
between swimmers having a narrower distribution. In both cases the 
deflection curve $\bar{\theta}_{\rm f}(\theta_\ell)$ is less steep as confirmed by the red curve 
  in Fig.~\ref{fig:deflection_numcomp} and the green curve in Fig.~\ref{fig:numstat_lightdeflection}.
  Therefore the strength of noise on the swimmer reorientations flattens the curve $\bar{\theta}_{\rm f}(\theta_\ell)$ 
  independent of the nature of the reorientations.


\section{Analytical modeling}
\label{sec:analytical}

\subsection{A  swimmer model in an anisotropic scattering medium} \label{sec:minimal:model}

We consider a simple theoretical model consisting of a self-propelled particle immersed in an effective anisotropic scattering medium \cite{brun2019effective}. The self-propelled particle is characterized by its 2D position ${\bf r}$ and an angle $\theta$ defining its direction of motion. The particle moves at a constant speed $v_0$. 
In the spirit of the numerical model studied in Sec.~\ref{sec:numerical}, 
we first neglect angular diffusion, and retain only the two main physical ingredients which 
are the scattering by the pillars and the random reorientations towards the direction of light.
To make the problem tractable, the lattice of pillars is modeled 
as an effective anisotropic scattering medium by following Ref.~\cite{brun2019effective}.
Guided by the numerical simulations of Sec.~\ref{Scattstat} (see Fig.~\ref{fig:numstat_scatter}), we choose a scattering rate
\be \label{eq:Ltheta0}
\lambda(\theta) = \lambda_0 [1- \cos(4\theta)] \,,
\ee
that depends on the orientation $\theta$ of the self-propelled particle.
Note that the scattering rate $\lambda(\theta)$ is not identical, but rather proportional to the scattering function defined in Sec.~\ref{Scattstat}. The proportionality factor is expected to be of the order of $v_0$ divided by the unit pillar cell size.
After a scattering event, the new angle $\theta'$ is randomly chosen from a uniform distribution over the interval $(-\pi,\pi]$.
The effective medium is homogeneous (though anisotropic), meaning that there 
are no explicit pillars, and scattering takes place with a probability rate $\lambda(\theta)$. 
It can thus occur at any place, and time intervals between stochastic scattering are exponentially distributed, 
with a mean value $1/\lambda(\theta)$.
The form Eq.~(\ref{eq:Ltheta0}) of the scattering rate implies that particles can travel freely, without being scattered, when their direction of motion is aligned either with the $x$ or $y$ axis.

In addition, we assume that the particle tends to reorient stochastically
during its motion to a direction, defined by an angle $\theta_\ell$ with the $x$-axis, opposite to the direction of the light source. For convenience, we call it below the direction of the light source, even though the swimmer actually moves away from the light source.
To be more specific, reorientation events occur with a probability $\alpha=1/\tau_{\rm ph}$ per unit time, and the new orientation $\theta'$ 
is chosen here exactly as the direction $\theta_{\ell}$ of the light source.
Using periodic boundary conditions, we also assume that the system reaches a spatially homogeneous state.
In this minimal model, the average direction of motion of the self-propelled particles can be computed exactly (see Appendix \ref{appendi1}), and it is found that the average angle $\bar{\theta}_{\rm f}$ within the scattering medium is equal to the angle $\theta_\ell$ defining the direction of the light source. Hence, there is on average no deflection by the scattering medium. This means that the present minimal model is not able to reproduce, even qualitatively, the deflection phenomenon observed in the experiment and in the numerics.
The physical ingredients that have been neglected here are notably the angular diffusion of the orientation of the self-propelled particle, and the angular fluctuations in the reorientation along the light direction. We will see below that taking into account these sources of noise is key to reproduce the phenomenology observed in the experiment and numerical simulations.

\subsection{Swimmer model in an anisotropic scattering medium with random reorientations}

We now slightly generalize the above model, by introducing angular diffusion in the motion of the swimmer, as well as some randomness in the angle chosen when reorienting in the light direction. We thus start by considering an 
active Brownian particle \cite{Cates:2013} such that in
the absence of scattering medium, the angle $\theta$ has a purely diffusive dynamics 
\be
\dot{\bf r} = v_0 {\bf e}(\theta) \,, \quad
\dot{\theta} = \xi(t)
\ee
where $\xi(t)$ is a white noise satisfying $\langle \xi(t) \rangle = 0$ and
\be
\langle \xi(t)\xi(t') \rangle = 2D_{\rm R} \, \delta(t-t') \,.
\ee
The angular diffusion coefficient is related to the persistence time $\tau$ by
$\tau=1/D_{\rm R}$.
In the presence of scattering medium, the angle $\theta$ is subjected as in the previous model to a random scattering with a rate $\lambda(\theta) = \lambda_0 - \lambda_0 \cos(4\theta)$, the angle $\theta'$ after scattering being uniformly distributed.
In addition, the reorientation process is also assumed to be noisy, in the sense that the angle $\theta'$ after reorientation is randomly chosen from a distribution
$\psi(\theta'-\theta_\ell)$ centered around the 
direction $\theta_\ell$ of the light source, similarly to the model used in \cite{martin2016photofocusing} that reproduces
the angular distribution of Fig.~\ref{orientation_simplemedium}. 
For simplicity, we assume that the distribution $\psi$ is symmetric, i.e., $\psi(-\theta)=\psi(\theta)$.

In this model, the deflection angle can no longer be computed exactly.
However, it can be evaluated using an approximation scheme, valid in a regime where the angular diffusion is not too small (i.e., $9D_{\rm R}\gtrsim \lambda_0+\alpha$). Under this approximation, we can evaluate the deflection angle $\phi$ defined as
\be \label{eq:thetaf:thetai}
\bar{\theta}_{\rm f} = \theta_\ell + \phi
\ee
through the relation
\be \label{eq:tan:phi}
\tan \phi = - \frac{\lambda_0 A_3 \psi_3 \sin(4\theta_\ell)}{2\psi_1+\lambda_0 A_3 \psi_3 \cos(4\theta_\ell)}\,.
\ee
Here, the notation $A_3$ denotes
\be
A_3 = \frac{1}{9 D_{\rm R} + \lambda_0 + \alpha},
\label{A3}
\ee
and $\psi_k$ is the Fourier coefficient of the distribution $\psi(\theta)$
in Fig.~\ref{orientation_simplemedium} obtained experimentally:
\be
\psi_k = \int_{-\pi}^{\pi} d\theta \, \psi(\theta) \,\cos(k\theta).
\label{psik}
\ee
The derivation of these results is reported in Appendix~\ref{appendi1}.
Consistently with the experimental results, the deviation $\phi$ vanishes when the angle $\theta_\ell$ of the light source is a multiple of $\frac{\pi}{2}$.
Before focusing on matching this theoretical model with the experimental results, let us briefly discuss the behavior of the deflection $\phi$ with the parameters of the model. We note that the angular diffusion as well as the width of the angular distribution $\psi(\theta)$ after reorientation, play a key role in determining the overall amplitude of the deflection. This result is consistent with the observation that the noise also plays an important role in determining the deflection in the numerical simulations reported in Sec.~\ref{sec:numerical}.
As mentioned above, the approximate expression  of $\phi$ in Eq.~(\ref{eq:tan:phi}) has been derived under 
the assumption $9D_{\rm R}\gtrsim \lambda_0+\alpha$.
Under this hypothesis, we see that the deflection $\phi$ decreases when increasing the angular 
diffusion coefficient $D_{\rm R}$. Similarly, increasing the width of the distribution $\psi(\theta)$ 
of the swimming angle $\theta$ after reorientation leads to a decrease of the deflection $\phi$,
in qualitative agreement with the results of numerical simulations displayed on Fig.~\ref{fig:numstat_lightdeflection}.
Hence we again observe that increasing the noise in the dynamics reduces the amplitude of 
the deflection (although, as noted above, at zero angular noise the deflection also vanishes). 
In other words, a finite amount of noise in the angular dynamics is needed to observe a deflection. 
This deflection disappears both for small (see Sec.~\ref{sec:minimal:model}) and large noise. 
This can be understood intuitively as follows. Angular diffusion actually allows the particle 
to explore orientations that are close to the direction $\theta_\ell$ of light. 
Through this local angular exploration, the particle can ``feel'' the anisotropy of the scattering 
rate $\lambda(\theta)$. If the particle has an angle $\theta$ slightly larger than $\theta_\ell$ ($0<\theta_\ell<\pi/4$),
it may for instance be scattered more than if it has an angle $\theta$ slightly 
smaller than $\theta_\ell$. In this case, this results in a slight deflection 
towards angles smaller than $\theta_\ell$.

We now turn to a comparison of the model with the experiment.
We provide in Fig.~\ref{angle_experimental} the mean deflected angle $\bar{\theta}_{\rm f}$ 
as a function of $\theta_\ell$ and compare it to our experimental results,
where $\theta_\ell$ is evaluated as the mean incidence angle $\bar{\theta}_{\rm i}$.
We fixed the parameters of the 
analytical fit of the experimental data using previously determined parameters: 
we fixed $\lambda_0=\unit{2.09}{\second}^{-1}$ associated with a medium where $d=\unit{30}{\micro\meter}$, 
the rotational diffusion coefficient $D_{\rm R}=\unit{0.37}{\second}^{-1}$ (both referenced in \cite{brun2019effective}), 
and the tumbling rate towards the light beam $\alpha=\unit{0.33}{\second}^{-1}$ \cite{martin2016photofocusing}. 
The distribution of orientations in Fig.~\ref{orientation_simplemedium} is shown to be well described 
by a truncated Lorentzian, providing a full width at half maximum of $\Gamma=0.436\,{\rm rad}=25\degree$. Then Eq.~(\ref{psik}) 
yields the values $\psi_1=0.809$ and $\psi_3=0.521$ for the Fourier coefficients of the distributions $\psi(\theta)$.
We obtain a quite good quantitative agreement with the experiments and the numerical simulations, demonstrating that we can explain 
this deflected phototactic swimming by means of a simple stochastic model. 
%

Note that in the numerical model studied in Sec.~\ref{sec:numerical}, no angular diffusion has been explicitly introduced. However, one may interpret the angular diffusion as an effective one emerging from the collisions with the pillars. In the analytical model, collisions with the pillar are modeled with a scattering rule where the angle after collision is randomly chosen in an isotropic way. Yet, in the numerical simulations, the angle after collision with a pillar is correlated with the angle before collision. A minimal way to account for this correlation is to introduce an effective angular diffusion that comes on top of the scattering rate.

\section{Discussion and Conclusion\label{sec:conclus}}

We investigated in the presence of a light stimulus the mean swimming direction
of a phototactic microalga {\it Chlamydomonas Reinhardtii},   described by the angle $\bar{\theta}_{\rm f}$,
through a square lattice of pillars with two symmetry axes along the $x$ and $y$ axes. 
We designed the experimental set up so that the distance between pillar centers is comparable to the persistence length of the swimmers, to allow for an interplay between the natural angular diffusion of swimmer motion and the scattering by the pillars.
We used in experiments a  CR  species that swims away from a light source. Their 
swimming direction is on the one hand guided by the direction of light, described by $\theta_\ell$, and
on the other hand  by the pillar symmetry axes. 
We found an interesting nonlinear $\theta_\ell$-dependence of 
the difference $\bar{\theta}_{\rm f}-\theta_\ell$. It vanishes for  the light beam parallel
to one of the symmetry axes (including the diagonal one) and this angle difference shows maxima with the light beam making an angle $\theta_\ell \sim 30\degree$ with the $x$ axis.

In order to further understand the origin of the nonlinear $\theta_\ell$-dependence of  
$\bar{\theta}_{\rm f}-\theta_\ell$, we complemented our experiments by simulations of a swimmer model and  by an analytical modeling
that contains the key ingredients leading to swimmer deflection in a square pillar lattice.

In our simulations we either neglected
or took into account the   hydrodynamic interactions (HI) between  pillars 
and swimmers. We found for single swimmers (diluted limit), that there is no qualitative  difference  
with and without HI. In simulations it is also possible to vary systematically the  
random reorientation distribution of swimmers, which is an intrinsic property of CR.
We found that this reorientation distribution has a strong 
impact on the $\theta_\ell$-dependence of $\bar{\theta}_{\rm f}-\theta_\ell$. For  a narrow reorientation distribution 
we found a strong channeling effect, i.e., with deviations of $\theta_\ell$ up to about $30^\degree$ from the $x$ or the $y$ axes
the swimmers moved essentially along the $x$ or the $y$ axes. 
This channeling effect is reduced by 
 choosing wider swimmer reorientation distributions. 
Thus, a narrow reorientation distribution of swimmers, which is associated with a strong channeling effect,  is the origin of a strong nonlinear 
nonlinear $\theta_\ell$-dependence of the difference $\bar{\theta}_{\rm f}-\theta_\ell$.

The reorientation distribution of single swimmers is specific to the chosen species. However, as we have shown by
numerical simulations,  it also depends on the hydrodynamic interaction between several swimmers.
If there are several swimmers in a pillar unit cell they  mutually influence via the 
nonlinear hydrodynamic interactions their individual dynamics and this HI acts like a broadening 
of their specific reorientation distribution. This broadening effect 
leads to a less pronounced nonlinear $\theta_\ell$ dependence of the
deflection $\bar{\theta}_{\rm f}-\theta_\ell$.
Therefore, the width of the reorientation distribution is a central
parameter that needs to be included in the models.

We also performed a numerical scattering statistics of a single model swimmer in the pillar lattice. 
A resulting scattering function in Fig.~\ref{fig:numstat_theta_s} is fitted by a phenomenological scattering rate given
by Eq.~(\ref{eq:Ltheta0}), that was introduced earlier with an analytical model of 
swimmer scattering in an anisotropic medium \cite{brun2019effective}, also described in Sec.~\ref{sec:minimal:model}.
Hence, with our numerical scattering statistics for single swimmers in the pillar lattice 
we found a `microscopic'  foundation for the phenomenological scattering rate used in the analytical model.

Along this reasoning we identified the  two essential parameters for the deflection of light guided swimmers
through  a square pillar lattice. This is one
phenomenological scattering rate 
and a parameter for the width of the random reorientation of swimmers, independent of its origin, which can be purely
intrinsic or a combination of intrinsic noise with HI effects.
This basic understanding of the swimmer deflection 
in a pillar lattice is condensed in our analytical model. Taking the two parameters of the basic model as fit parameters
we obtain a very good 
agreement with the experimental results. 
A great advantage of our  analytical model is its simple handing and 
is also appropriate for different pillar
sizes and distances through an associated effective scattering medium.  It can also be adapted to different geometries (triangular, hexagonal,...) 
by taking into account the appropriate angular dependence of $\lambda(\theta)$. 

The identified dependence of the swimmer deflection on the 
reorientation distribution of swimmers in a pillar lattice may be also used for separating 
swimmers with a different reorientation distribution by guiding them via a light source through a pillar lattice.

\section*{Acknowledgements}
For support, we thank the Franco-German University (Grant No. CFDA-Q1-14, ''Living Fluids'').


\appendix


\section{Numerical method}
\label{appendi2}

\subsection{Swimmer model}
\label{appswimmmodel}
\begin{figure}[!ht]
\includegraphics[width=7 cm]{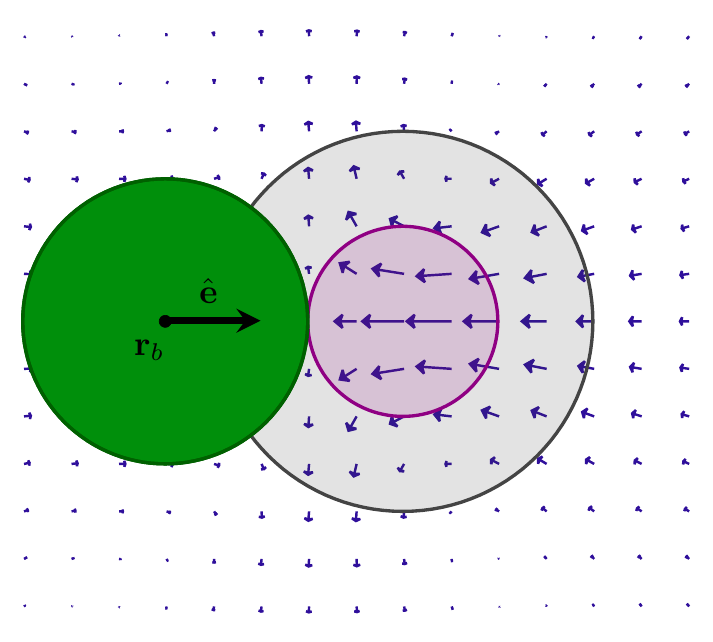}
\caption{
	Numerical representation of a CR at position $\mathbf{r}$ and swimming direction $\hat{\mathbf{e}}$. The driving force
	$\mathbf{f}_{\rm d}$ acting on the swimmer body is balanced by applying a negative force 
	$-\mathbf{f}_{\rm d}$ at the center of the
	flagella moving range onto the fluid (purple circle). The blue arrows show the resulting flow-field for simulations with
	the LB method, which resembles the time-averaged field of a Chlamydomonas \cite{Drescher:2010}.}
\label{fig:sim_model_app}
\end{figure}

The swimmer is modeled as a force-dipole swimmer (Fig.~\ref{fig:sim_model_app}) with the orientation $\hat{\mathbf{e}}$.  
For the force calculation, we consider excluded volume forces $\mathbf{f}_{\rm e}(\mathbf{r})$ for both, the flagella and the body. Here, $\mathbf{f}_{\rm e}(\mathbf{r})$ is the short-range repulsive part of the Weeks-Chandler-Anderson potential \cite{Weeks:1971} for a separation distance $\mathbf{r}$. The driving force $\mathbf{f}_{\rm d}=f_{\rm d}\,\hat{\mathbf{e}}$ acting on the body is balanced by applying a negative force $-\mathbf{f}_{\rm d}$ on the fluid at the position of the flagella 
(see Fig.~\ref{fig:sim_model_app}). We assume that the flagella have negligible mass and the center of mass is located at the center of the body, resulting in a torque due to the excluded volume force with center at $\mathbf{r}_{\rm f}$.  The total, non-hydrodynamic forces $\mathbf{f}_{\rm b}$ and torques $\mathbf{t}_{\rm b}$ acting on the swimmer body are given by
\bea
    \mathbf{f}_{\rm b} &=& \mathbf{f}_{\rm d} + \sum_i \left[ \mathbf{f}_{\rm e} (\mathbf{e}_i - \mathbf{r}_{\rm b}) + \mathbf{f}_{\rm e} (\mathbf{e}_i - \mathbf{r}_{\rm f}) \right],  \\
    \mathbf{t}_{\rm b} &=& \sum_i \left(\mathbf{r}_{\rm f}-\mathbf{r}_{\rm b}\right) \times \mathbf{f}_{\rm e} (\mathbf{e}_i - \mathbf{r}_{\rm f}),
\eea
where we sum over all other objects $i$ at positions $\mathbf{e}_i$ to be considered for excluded volume interactions ({\it i.e.} other swimmers and obstacles). 

\subsection{Equations of motion}
\label{sec:swimmmereq}
As fluid solver we use two different models, the dissipative collision model (DCM) and the Lattice-Boltzmann (LB) method. Both require as input the positions, forces and torques $\mathbf{r}_{\rm b},\mathbf{f}_{\rm b},\mathbf{t}_{\rm b}$ of the body and the positions $\mathbf{r}_{\rm f}$ and forces $\mathbf{f}_{\rm f}=-\mathbf{f}_{\rm d}$ at the flagella position at time $t$. As output, they provide the velocities $\mathbf{v}_{\rm b}$ and angular velocities $\mathbf{w}_{\rm b}$ for the new time $t+\Delta t$. The new positions and swimming-directions are then updated via an Euler's integration: 
\bea
\mathbf{r}_{\rm b}(t+\Delta t)&=&\mathbf{r}_{\rm b}(t)+\Delta t \, \mathbf{v}_{\rm b}(t), \\
\hat{\mathbf{e}}(t+\Delta t)&=& \mathbf{R}(\mathbf{w}_{\rm b}(t)\, \Delta t)\, \hat{\mathbf{e}},
\eea
where $\mathbf{R}({\bm \alpha})$ is the rotation defined by the vector ${\bm \alpha}$.
\paragraph{Dissipative collision model (DCM):}
For the dissipative collision model, we only consider the driving force and collisions through excluded volume interactions while neglecting HI. We use the Stokes-drag of the particles to calculate the new angular and translational velocities as
\bea
\mathbf{v}_{\rm b}(t+\Delta t) = \frac{\mathbf{f}_{\rm b}}{6 \pi \eta a}, \\
\mathbf{w}_{\rm b}(t+\Delta t) = \frac{\mathbf{t}_{\rm b}}{8 \pi \eta a^3}.
\eea
For the simulations, we use as parameters $\eta=1/6$ and a radius of $a=3$. 
\paragraph{Lattice Boltzmann (LB) method:} For the simulations including HI, we utilize the LB method with the Bhatnagar-Gross-Krook (BGK) collision step which reproduces the full 
Navier-Stokes equation in the incompressible limit \cite{Aidun:2010}. We calculate the phase-density $f_i(\mathbf{x},t)$ of the fluid elements on a three-dimensional grid of positions $\mathbf{x}=(x,y,z)$ along the discrete directions $\mathbf{c}_i (i=0,\dots,18)$ (D3Q19 model) with a spatial discretization of $\Delta x=1$ and $\Delta t=1$ for the temporal discretization. 
The evolution equation is given by  \cite{Bhatnagar:1954,Aidun:2010}
\be
\label{eq:sim_bgk}
f_i(\mathbf{x}+\mathbf{c}_i\,\Delta t,t+\Delta t) = f_i(\mathbf{x},t)  + \mathcal{C}
\ee
where
\be
\mathcal{C}=\frac{1}{\tau} \left[ f_i (\mathbf{x},t) - f^{\rm eq}_i(\mathbf{x},t) \right]
\ee
is the BGK collision operator with the equilibrium distribution
\be
f^{\rm eq}_i (\mbf{x},t)= \rho w_i  \left[ 1 + \frac{\mbf{c}_i \cdot \mbf{u}}{c_{\rm s}^2} + \frac{ (\mbf{c}_i \cdot \mbf{u})^2}{2 c_{\rm s}^4} - \frac{\mbf{u}^2}{2 c_{\rm s}^2}\right].
\ee
The time constant $\tau$ is linked to the fluid viscosity via $\nu=c_{\rm s}^2 \Delta t (\tau-1/2)$. The weighting factors $w_i$ and 
the parameter $c_{\rm s}$ are constants with specific values for the chosen simulation model
\cite{Aidun:2010}. Walls are implemented with the standard bounce-back (bbk) scheme \cite{Ladd:1994:1}, 
which alters the evolution equation (\ref{eq:sim_bgk}) such as:
\be
f_{i^\prime}(\mathbf{x}, t+\Delta t) = f_i(\mathbf{x},t)  + \mathcal{C} + \mathcal{W},
\ee
if $f_i$ points into a wall, 
where $i^\prime$ is the antiparallel direction to 
$i$ and $\mathcal{W}=2w_i \rho \frac{\mbf{c}_i \cdot \mbf{u}_{\rm w}}{c_{\rm s}^2}$ 
accounts for the momentum exchange of a moving wall with velocity $\mbf{u}_{\rm w}$.

External volume forces linked to the out-of-lattice position of the swimmers are coupled to the fluid grid via the immersed boundary method using the four-point stencil \cite{Peskin:2002}, which provides the volume-forces $\mbf{F}_{\rm v} (\mbf{x})$ at the fluid grid positions. For nodes with $\mbf{F}_{\rm v}\neq 0$, the collision operator
[Eq.~(\ref{eq:sim_bgk})] is extended by adding the Guo force-coupling term \cite{Guo:2002}
\be
\mathcal{F}=\Delta t \left(1-\frac{1}{2\tau}\right) w_i \left[ \frac{\mbf{c}_i-\mbf{u}}{c_{\rm s}^2} + \frac{(\mbf{c}_i\cdot \mbf{u})}{c_{\rm s}^4} \mbf{c}_i\right] \cdot \mathbf{F}_{\rm v}.
\ee
The fluid-density $\rho$ and fluid-velocity $\mbf{u}$ are obtained by 
\bea
\rho &=& \sum_i f_i ,\\
\rho \mbf{u}&=&\sum c_i\mbf{f}_i + \frac{\Delta t}{2} \mbf{F}_{\rm v}.
\eea
The swimmers body is implemented by setting links crossing the particle surface as moving-wall \cite{Aidun:1995,Aidun:1998}. The wall velocity for those links is set to  $\mbf{u}_{\rm w}(\mbf{x})=\mbf{v}_{\rm b}+\mbf{w}\times(\mbf{x}+\frac{1}{2}\mbf{c}_i-\mbf{r}_{\rm b}) $. The hydrodynamic force $\mbf{f}_{\rm h}$ and torque $\mbf{t}_{\rm h}$ exerted from the fluid on the particle can then be calculated by summing all contributions of the momentum exchange between the fluid and wall-links over the surface of the body and eventual covered/uncovered forces (see \cite{Aidun:1998} for more details). 
The new swimmer velocity is then given by Newton's law according to
\bea
\mathbf{v}_{\rm b}(t+\Delta t) &=& \mathbf{v}_{\rm b}(t) + \frac{\Delta t}{M}
\left[\overline{\mathbf{f}}_{\rm h} + \overline{\mathbf{f}_{\rm b}}\right]  ,\\
\mathbf{w}_{\rm b}(t+\Delta t) &=& \mathbf{w}_{\rm b}(t) + \frac{\Delta t}{I}
\left[\overline{\mathbf{t}}_{\rm h} + \overline{\mathbf{t}_{\rm b}}\right],
\eea
where $M$ is the mass of the swimmer body and $I$ its moment of inertia while $\overline{\mathbf{f}}_{\rm h}$ and $\overline{\mathbf{t}}_{\rm h}$ are the forces and torques averaged over two intermediate time-steps as described in \cite{Aidun:1998}.\\
For the swimmer, we use the radius $a=3$, the density $\rho=1$ and a relaxation parameter $\tau=1$. For the dipole-force, we choose $f_{\rm d}=0.25$ by using the technique described in \cite{Cates:2004} to find a reasonable swimming velocity which ensures low Reynolds dynamics while keeping simulation time short. This parameter results in a Reynolds-number of $0.39$ and an error of less than two percent in a distance of one radius away from the Chlamydomonas.


\section{Analytical model}
\label{appendi1}

We provide in this Appendix a detailed analysis of the effective medium model defined in Sec.~\ref{sec:analytical} of the main text, in order to evaluate analytically the deflection angle.

Assuming for simplicity spatial homogeneity, the dynamical distribution $P(\theta)$ of the swimmer angle $\theta$ satisfies the evolution equation
\bea \label{eq:dyn:Ptheta}
\partial_t P(\theta) &=& D_{\rm R} \partial_{\theta}^2 P(\theta) - \big( \lambda(\theta)+ \alpha\big) P(\theta)\\ \nonumber
&&
+ \frac{1}{2\pi} \int_{-\pi}^{\pi} d\theta' \, \lambda(\theta') P(\theta')
+ \alpha \rho \psi(\theta-\theta_\ell)\,.
\eea
We assume in the following that $P(\theta)$ is normalized as $\int_{-\pi}^{\pi}d\theta\, P(\theta)=\rho$, where $\rho$ is the uniform density of swimmers.
It is convenient to define the angular Fourier mode $f_k$ of the distribution $P(\theta)$, as
\be
f_k = \int_{-\pi}^{\pi}d\theta\, P(\theta) \, e^{ik\theta}\,.
\ee
Note that $f_{-k}=f_k^*$, where the star indicates the complex conjugate.
Expanding Eq.~(\ref{eq:dyn:Ptheta}) in Fourier modes, one gets for 
$k\neq 0$ (the equation for $k=0$ is trivially valid in a spatially homogeneous state),
\bea \label{eq:Fourier:fk}
\partial_t f_k &= -(D_{\rm R} k^2 + \lambda_0 + \alpha) f_k
+\alpha \rho \psi_k e^{ik\theta_\ell} \nonumber \\
& \qquad + \frac{\lambda_0}{2} (f_{k+4}+f_{k-4})
\eea
where $\psi_k$ is the Fourier coefficient of the distribution $\psi(\theta)$ of the angle $\theta$ after 
random reorientation to the light source [see Eq.~(\ref{psik}) of the main text], and $\theta_\ell$ 
is the average direction towards which the particle reorients.
We wish to determine the average velocity ${\bf v}$ of particles in the presence of the light source. The velocity ${\bf v}$ is related to the Fourier mode $f_1$ through
\be \label{eq:v:rho}
{\bf v} = \frac{v_0}{\rho} \, ({\rm Re} f_1, {\rm Im}f_1)\,.
\ee
The average angle of motion of the microswimmers in the effective medium, corresponding to the direction of the velocity ${\bf v}$ is thus given by
\be \label{eq:thetaf:argf1}
\bar{\theta}_{\rm f} = {\rm Arg}(f_1)
\ee
where the function ${\rm Arg}(z)$ is the argument of the complex number $z$.
We thus need to determine $f_1$ in the stationary homogeneous state. 
Dropping the time derivative term in Eq.~(\ref{eq:Fourier:fk}),
one has to solve the infinite hierarchy of equations
\be
\label{eq:fk:light}
f_k = A_k \left[ \alpha \rho \psi_k e^{ik\theta_\ell} + \frac{\lambda_0}{2}(f_{k+4}+f_{k-4}) \right]
\ee
where the parameter $A_k$ is defined as
\be
A_k = \frac{1}{k^2 D_{\rm R} + \lambda_0 + \alpha} \,.
\ee
In general, the hierarchy of equations (\ref{eq:fk:light}) cannot be solved exactly, at least not in a simple way.
However, in the specific case $D_{\rm R}=0$ (absence of angular diffusion) and $\psi_k=1$ for all $k$, corresponding to a Dirac distribution $\psi(\theta)$, an exact solution can be found because Eq.~(\ref{eq:fk:light}) becomes in this limit a simple recursion relation for the Fourier modes $f_{4n+1}$ (other modes do not need to be considered). Solving this recursion relation to determine $f_{4n+1}$ for all $n$, one eventually finds for $f_1$,
\be
f_1 = \frac{\rho \alpha e^{i\theta_\ell}}{\lambda_0+\alpha-\lambda_0\cos(4\bar{\theta}_{\rm i})}\,,
\ee
which implies $\bar{\theta}_{\rm f}=\theta_\ell$. Hence in the absence of angular noise and with an infinitely sharp distribution $\psi(\theta)$, there is no deflection in the effective medium model.

In other cases, a simple solution cannot be found, and one has to resort to an approximation scheme. We discuss below a simple approximation scheme, together with its range of validity.
We first note that if $D_{\rm R}$ is not too small as compared to $\lambda_0+\alpha$, the coefficient $A_k$ decays relatively rapidly when $k$ is increased.
A simple approximation scheme is thus to approximate $A_k$ by zero beyond some order $k$.
Using the previously reported values of $D_{\rm R}$, $\lambda_0$ and $\alpha$ \cite{brun2019effective,martin2016photofocusing} (see also Sec.~\ref{sec:analytical}), we find that for $k>4$ the term $k^2D_{\rm R}$ starts to be dominant over $\lambda_0+\alpha$, thus making $A_k$ decay faster for higher values of $k$.
We thus make the crude approximation $A_k \approx 0$ for $k>4$.
From Eq.~(\ref{eq:fk:light}), this implies that one can neglect Fourier modes $f_k$ with $|k|>4$, leading to the following equations for $f_1$ and $f_3$,
\bea
\label{eq:f1:light}
f_1 &=& A_1 \left[ \alpha \rho \psi_1 e^{i\theta_\ell} + \frac{\lambda_0}{2} f_3^* \right] ,\\
\label{eq:f3:light}
f_3 &=& A_3 \left[ \alpha \rho \psi_3 e^{3i\theta_\ell} + \frac{\lambda_0}{2} f_1^* \right].
\eea
Combining Eqs.~(\ref{eq:f1:light}) and (\ref{eq:f3:light}), one obtains
\be \label{eq:f1:final}
f_1 = \frac{\alpha A_1 \rho}{1-\frac{\lambda_0^2}{4} A_1 A_3}
\left[\psi_1 + \frac{\lambda_0}{2}{A_3 \psi_3} e^{-4i\theta_\ell} \right]
e^{i\theta_\ell}\,.
\ee
The prefactor in front of the bracket is always positive, and taking the argument of Eq.~(\ref{eq:f1:final}) to evaluate $\bar{\theta}_{\rm f}$ according to Eq.~(\ref{eq:thetaf:argf1}), one finds
\be \label{eq:thetaf:thetai:app}
\bar{\theta}_{\rm f} = \theta_\ell + \phi
\ee
where the deflection angle $\phi$ is determined by
\be \label{eq:tan:phi:app}
\tan \phi = - \frac{\lambda_0 A_3 \psi_3 \sin(4\theta_\ell)}{2\psi_1+\lambda_0 A_3 \psi_3 \cos(4\theta_\ell)} \,.
\ee
Note that the deflection $\phi=0$ when $\theta_\ell$ is a multiple of $\frac{\pi}{2}$. 
According to the approximations made, the expression (\ref{eq:tan:phi:app}) of the deflection angle is expected to be approximately valid for not too small angular diffusion coefficient $D_{\rm R}$, that is, as long as $9D_{\rm R} \gtrsim \lambda_0+\alpha$, which is the case with the experimental values.
For smaller $D_{\rm R}$, the simple truncation procedure used above is no longer valid, and a larger number of Fourier modes should be retained in the approximation. 
Finding in an analytical way the approximate solution of the hierarchy of equations (\ref{eq:fk:light}) is thus more difficult, and one would then need to resort to a numerical procedure to solve Eq.~(\ref{eq:fk:light}).
The above results are summarized in Sec.~\ref{sec:analytical} of the main text.



%

\end{document}